\documentclass[twocolumn,preprintnumbers,amsmath,amssymb, prl]{revtex4}
\usepackage{graphicx}
\usepackage[tight]{subfigure}
\usepackage{graphicx}
\usepackage{subfigure}

\begin{document}

\title{Secondary structure of Ac-Ala$_n$-LysH$^+$ polyalanine peptides ($n$=5,10,15) in vacuo: Helical or not?}

\author{M. Rossi$^1$, V. Blum$^1$, P. Kupser$^1$, G. von Helden$^1$, F. Bierau$^1$,
  K. Pagel$^{2,\ast}$, G. Meijer$^1$, and M. Scheffler$^1$}
\affiliation{$^1$Fritz-Haber-Institut der Max-Planck-Gesellschaft, D-14195, Berlin, Germany}
\affiliation{$^2$Freie Universit\"at Berlin, Institut f\"ur Chemie und
  Biochemie,Takustr. 3, 14195 Berlin, Germany} 
\altaffiliation{University of Oxford, Chemistry Research Laboratory, Mansfield Road,
     Oxford OX1 3TA, UK}


\begin{abstract}
The polyalanine-based peptide series Ac-Ala$_n$-LysH$^+$ ($n$=5-20)
is a prime example that a secondary structure motif which is
well-known from the solution phase (here: helices) can be formed
\emph{in vacuo}. We here revisit this  
conclusion for $n$=5,10,15, using density-functional theory 
(van der Waals corrected generalized gradient approximation),
and gas-phase infrared vibrational spectroscopy. For the  
longer molecules ($n$=10,15) $\alpha$-helical models provide good
qualitative agreement (theory vs. experiment) already in the harmonic
approximation. For $n$=5, the lowest energy
conformer is not a simple helix, but competes closely with
$\alpha$-helical motifs at 300~K. Close agreement between  
infrared spectra from experiment and \emph{ab initio} molecular
dynamics (including anharmonic effects) supports our findings.
\end{abstract}

\maketitle

It is often said that the structure of peptides and
proteins can not be understood without the action of a solvent, and
this statement is certainly true for the full three-dimensional
(tertiary) structure of proteins. However, their \emph{secondary}
structure level (helices, sheets, turns) is predominantly
shaped by \emph{intramolecular} interactions---most importantly,
hydrogen bonds. For these interactions, benchmark experiment-theory
comparisons under well-defined ``clean-room'' conditions \emph{in
  vacuo} can furnish critical information towards a complete,
predictive picture of peptide structure and dynamics. For example,
first-principles approaches such as density-functional theory (DFT)
with popular exchange-correlation functionals do not account for van
der Waals (vdW) interactions. For peptide 
studies, precise experimental calibration points for different,
actively developed theoretical remedies \cite{wu-yang2002, langreth-lundqvist, jurecka06,
  grimme07, tkatchenko-scheffler} would be extremely useful---\emph{if
the same structure as in the solution phase can be formed}.

In a seminal ion-mobility spectrometry study more than a decade ago,
Hudgins, Ratner, and Jarrold (HRJ) 
\cite{hudgins-jarrold} reported the formation of just such a
secondary structure motif known from the solution phase (helical) \emph{in vacuo} for
a series of designed, charged polyalanine-based peptides
Ac-Ala$_n$-LysH$^+$ ($n$=5-20). While much follow-up work has been
done after the original HRJ study (e.g.,
Refs. \cite{hudgins-jarrold99, kohtani-jarrold2000,
  kohtani-jarrold2004, kohtani-jarrold2004b, stearns-rizzo2007,
  stearns-rizzo2009, vaden-paizs, cimas-vaden-gaigeot}) the 
``helical'' nature of the exact series Ac-Ala$_n$-LysH$^+$ 
is so far still established only indirectly by comparing ion-mobility
cross-sections to results from force-field based molecular
dynamics. A helical assignment for polyalanine is thus plausible
and easily accepted, but different structural conclusions are not
entirely ruled out. This fact is strikingly evidenced by additional ion-mobility 
experiments with microsolvation \cite{kohtani-jarrold2004},
which indicate that the secondary structure is 
not yet helical for $n<8$. On the other hand, a spectroscopic study of
Ac-Phe-Ala$_5$-LysH$^+$ \cite{stearns-rizzo2009} inferred structures
with ``helical'' H-bond rings ($\alpha$- or 3$_{10}$-helix-like) as the
dominant conformers. 

The key goal of the present work is to unambiguously verify the
structure of the $n$=5, 10, and 15 members of the original HRJ series,
both experimentally and theoretically. This is an important task, as
\emph{safely} knowing the correct structure is a 
key prerequisite for any further physical conclusions. Obviously, this
is a broadly important statement for a wide part of physics and chemistry ---surface
science and catalysis, alloys and compounds, semiconductor properties
etc.---but for a benchmark system such as the HRJ series,
such safe knowledge is
particularly crucial. On the theory side, we employ DFT in 
the PBE \cite{PBE} generalized gradient approximation corrected for vdW
interactions\cite{tkatchenko-scheffler} with an accuracy that is
critical for the success of our work. We thereby confirm the helical
assignment for $n$=10 and 15, but for $n$=5, the lowest energy
structure is indeed not a simple helix. Even for such a relatively
short molecule, there is then an enormous structural variety to
navigate, an effort which one must not shun (see below). We verify our findings against experimental 
infrared multiple photon dissociation (IRMPD) spectra of the
vibrational modes in the 1000-2000~cm$^{-1}$ region, which pertain to
finite $T\approx$300~K. Here, harmonic free-energy calculations show
that multiple conformers for $n$=5 (both helical and non-helical)
should coexist, and are supported by calculated vibrational 
spectra (harmonic and anharmonic) in close agreement with experiment.

For the experiments, the peptides were synthesized by standard Fmoc
chemistry. The experimental IR spectra were recorded using the
Fourier transform ion cyclotron (FT-ICR) mass spectrometer
\cite{valle-blakney} at the free-electron laser FELIX
\cite{oepts-amersfoort}. Ions were brought into the gas-phase by
electrospray ionization (ESI) ($\sim 1$ mg of peptide in 900 $\mu$l
TFA/100 $\mu$l H$_2$O) and mass selected and trapped inside the ICR
cell which is optically accessible. When the IR light
is resonant with an IR active vibrational mode of the molecule, many
photons can be absorbed, causing the dissociation of the ion
(IRMPD). Mass spectra are recorded after 4s of IR irradiation. Monitoring the depletion of the parent ion
signal and/or the fragmentation yield as a function of IR frequency leads to an IR spectrum. 

All DFT+vdW calculations for this work were performed using the FHI-aims
\cite{FHI-aims} program package for an accurate, all-electron
description based on numeric atom-centered orbitals. 
``Tight'' computational settings and accurate \emph{tier 2} basis sets
\cite{FHI-aims} were employed throughout. 
Harmonic vibrational
frequencies, intensities and free energies were computed from finite
differences. For Ac-Ala$_5$-LysH$^+$, we computed infrared intensities
$I(\omega)$ \emph{beyond} the harmonic approximation from
\emph{ab initio} molecular dynamics (AIMD) runs $>$20~ps ($NVE$
ensemble, with a 300~K $NVT$ equilibration), by calculating
the Fourier transform of the dipole auto-correlation function
\cite{gaigeot-vuilleumier, cimas-vaden-gaigeot, li-iyengar}
with a quantum corrector factor to the classical line
shape \cite{borysow-fromm, ramirez-marx} proportional to $\omega^2$
(see Ref. \cite{gaigeot-vuilleumier}). 
For a direct
comparison to experiment, it is important to represent the ``density
of states'' like nature of the measured spectra also in the calculated
curves. All calculated spectra (harmonic and anharmonic) are therefore
convoluted with a Gaussian broadening function with a variable width
of 1\% of the corresponding wave number, which accounts for the spectral
width of the excitation laser. Further broadening
mechanisms, e.g., due to the excitation process, are reflected in the
experimental data.\cite{Oomens06}. 

For the DFT+vdW calculations on Ac-Ala$_5$-LysH$^+$, we generate
a large body of possible starting conformations using the empirical
OPLS-AA force field (as given in Ref. \cite{oplsaa-2001} and
references therein) in a series of basin hopping 
structure searches performed with the TINKER \cite{tinker}
package. Our particular choice of force field
was not motivated by any other reason than that an input structure
``generator'' for DFT was needed. That said, the performance of OPLS-AA for 
gas-phase Alanine dipeptides and tetrapeptides was assessed rather
favorably in earlier benchmark
work.\cite{Beachy97,kaminski-jorgensen} In the searches, specific
constraints on one or more hydrogen bonds could 
be enforced. In total, we collected $O(10^5)$ nominally different
conformers from (i) an unconstrained search, (ii) one hydrogen bond in
the Ala$_5$ part constrained to remain $\alpha$-helical, (iii) two
hydrogen bonds in Ala$_5$ constrained to form a $3_{10}$-helix, (iv)
three hydrogen bonds in the Ala$_5$ part constrained to form a $2_7$
helix, or (v) one hydrogen bond in the full peptide constrained to a
$\pi$-helical form. 
As is well known~\cite{penev-ireta-shea}, conformational
energy differences between different types of secondary structure may
vary strongly between different force fields and/or DFT.
We reduce our reliance on the energy hierarchy provided by the
force field by following up with full DFT+vdW relaxations for a wide
range of conformers, 134 in total. This range includes the lowest
$\sim$0.3~eV for the unconstrained and $\alpha$-helical 
searches, and the lowest $\sim$0.15~eV for the 3$_{10}$-constrained
search, as well as the lowest few $\pi$- and 2$_7$-helical
candidates. Almost all $\pi$-helical geometries found 
in the force-field relaxed with DFT either into $\alpha$ or 3$_{10}$
helices, and all relaxed 2$_7$ helices were higher in
energy than our lowest-energy conformer by at least 0.26~eV. 

\begin{figure}[ht]
\begin{center}
\includegraphics[width=0.3\textwidth]{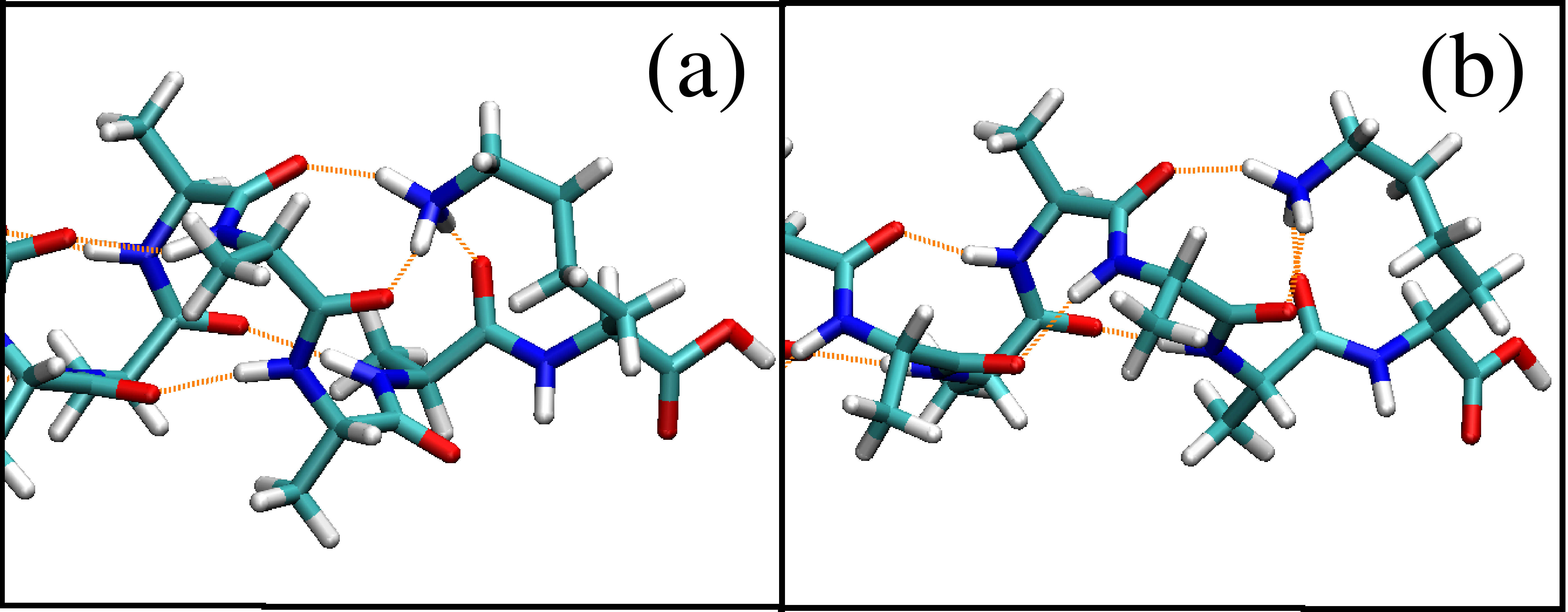} \\
\includegraphics[width=0.48\textwidth]{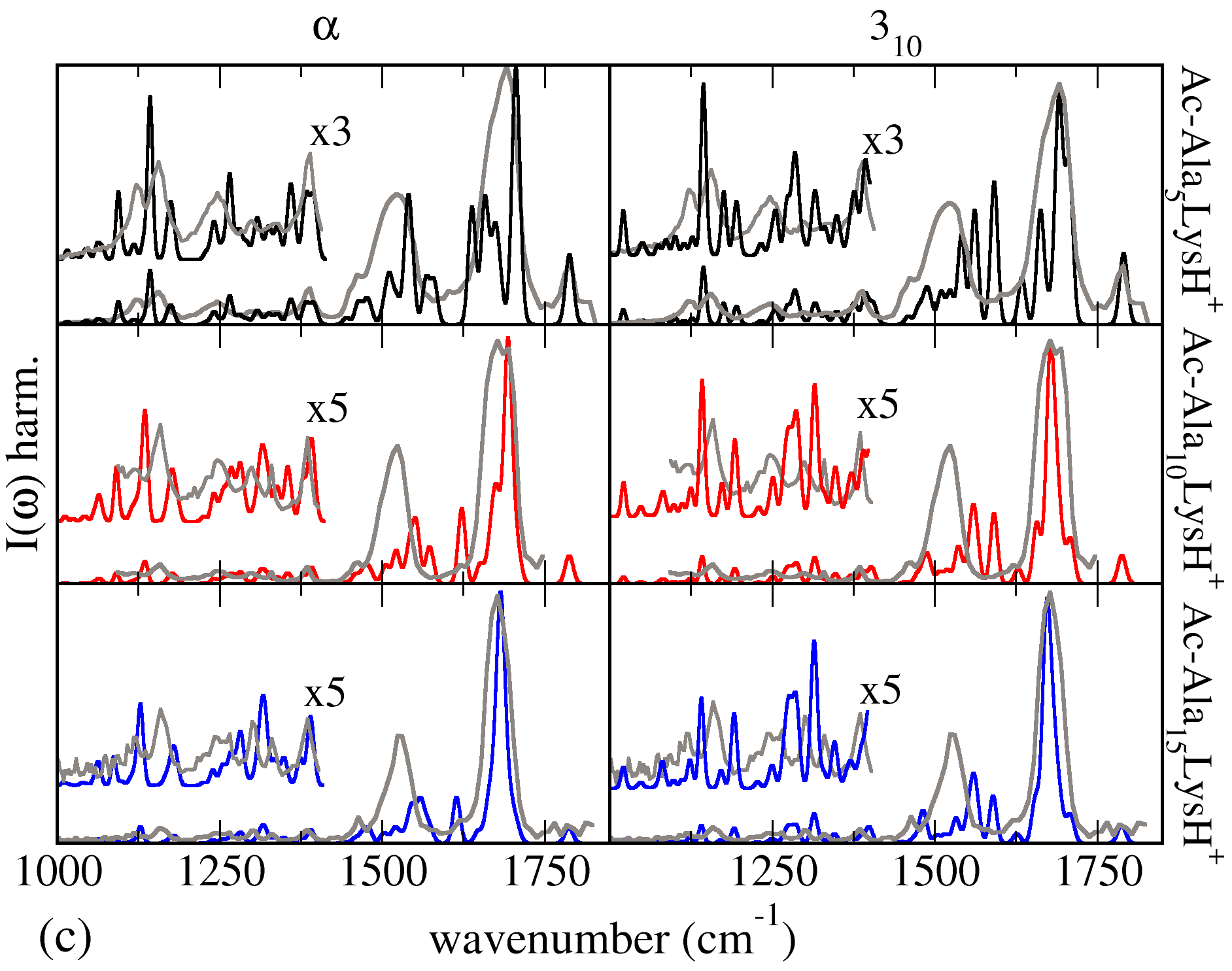}
\end{center}
\caption{\label{ala5-ala10-ala15}
Comparison between experimental [gray full lines in (c)] and calculated, broadened
harmonic vibrational spectra [black, red, and blue solid lines in (c)] for
$\alpha$-helices of Ac-Ala$_{5}$-LysH$^+$(top), Ac-Ala$_{10}$-LysH$^+$
(middle) and Ac-Ala$_{15}$-LysH$^+$ (bottom). Also shown are the
terminating H-bond networks used for $\alpha$ (a) and 3$_{10}$
(b). All theoretical results for $\alpha$- and 3$_{10}$-helices have been aligned to the free C-O peak at $\sim1790$cm$^{-1}$.}
\end{figure}

Figure \ref{ala5-ala10-ala15} shows the experimental IRMPD spectra for
the three lengths of peptides studied ($n$=5,10,15) and calculated vibrational spectra in the harmonic
approximation for two specific types of hydrogen bond networks:
$\alpha$-helical (left) and 3$_{10}$-helical (right). It is well known
that both the choice of the density functional and the neglect of
anharmonic effects will lead to characteristic frequency shifts
between theoretical and experimental spectra. For a better visual
comparison, all calculated spectra in the present work are therefore
rigidly shifted to be aligned with the approximate location of the
localized free C-O vibration at $\sim1790$cm$^{-1}$ in
experiment for $n$=5. For example, this shift amounts to $\approx$20~cm$^{-1}$
for $\alpha$-helical conformers. All intensities were uniformly scaled
to match the highest peak (Amide-I), but no further scaling factors
(frequency or intensity) were employed. Given the limitations of the
$T$=0 harmonic approximation when  
comparing to room-temperature experimental spectra, the agreement is
rather reasonable for the $\alpha$-helical conformers, while this is
much less the case in terms of relative peak positions and fine structure for the 3$_{10}$
helical conformers of $n$=10, 15. This observation correlates with
calculated energy differences, where the 3$_{10}$ helical conformers
are higher in energy than the $\alpha$-helical ones by 0.41~eV
($n$=10) and 0.82~eV ($n$=15). In addition, an OPLS-AA based 
basin-hopping structure enumeration for $n$=10 did not reveal any
non-$\alpha$ conformers within at least 0.15~eV. For $n$=15, the
employed structure search procedure becomes prohibitive, although
direct AIMD simulations show that $\alpha$-helix conformations are
structurally stable even at high $T$ (500~K) for at least several tens
of ps \cite{Tkatchenko-Rossi-Blum-Scheffler}. The available evidence
thus points to at least predominantly $\alpha$-helical secondary
structure for $n$=10, 15. In any case, the observed disagreements for
3$_{10}$ confirm the basic structure sensitivity of the measured IRMPD
spectra. 

\begin{figure}[ht]
\begin{center}
\includegraphics[width=0.48\textwidth]{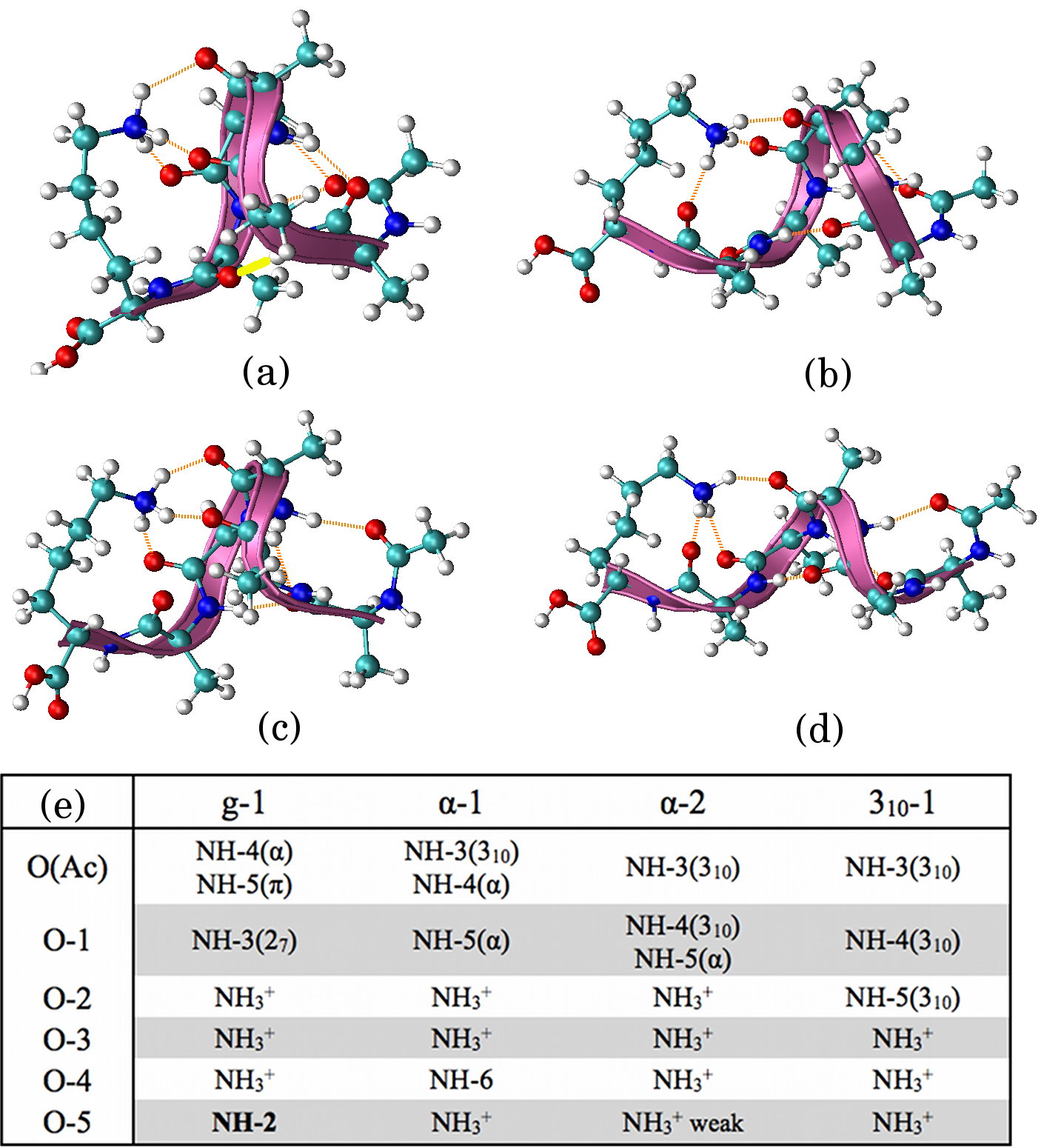}
\end{center}
\caption{Geometries of Ac-Ala$_5$-LysH$^+$: (a) g-1;
  (b) $\alpha$-1; (c) $\alpha$-2; (d) 3$_{10}$-1; and (e) H-bond network of the conformers: (C-)O and N-H groups are numbered 
  starting from the N terminus and ending at the C terminus.} 
\label{schematic-conformers}
\end{figure}

For Ac-Ala$_5$-LysH$^+$, the four lowest-energy conformers from our search and their H-bond
networks are shown in Fig. \ref{schematic-conformers}. Three of these
conformers (labeled $\alpha$-1, $\alpha$-2, 3$_{10}$-1) are 
``helical'', in the sense that they contain two
well-separated terminations with the appropriate $\alpha$- or
3$_{10}$-like H-bond loops in their Ala$_5$ section. The
lowest-energy conformer, labelled g-1, contains only one 2$_7$ like
loop, H-bonds to the NH$_3^+$ end of the Lys side chain, and one
H-bond that runs \emph{against} the normal helix dipole, effectively
short-circuiting the terminations. In fact, small structural
differences in the Lys side chain lead to three nonequivalent
conformers with the g-1 H-bond network, only one of which is shown
here for simplicity. 

\begin{figure}[ht]
\begin{center}
\includegraphics[width=0.47\textwidth]{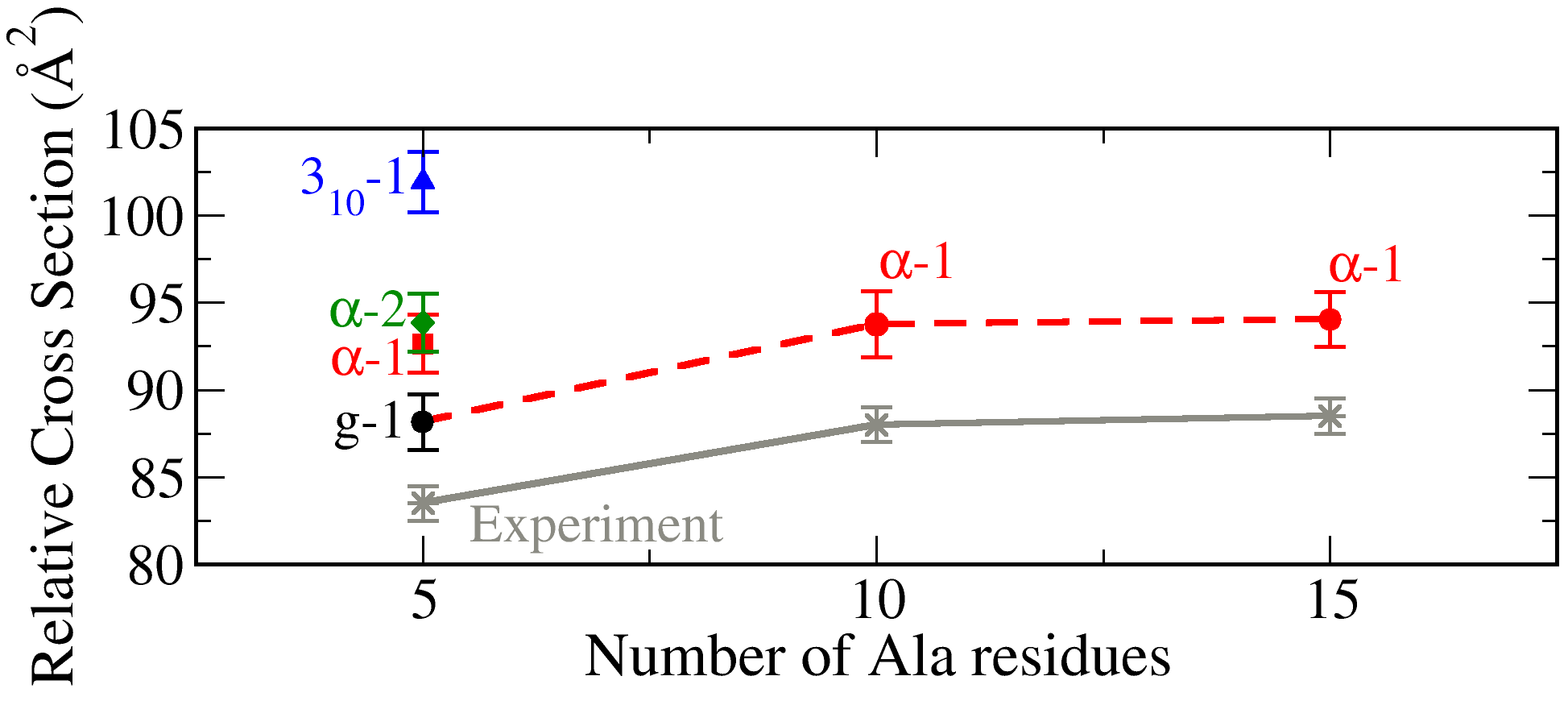}
\end{center}
\caption{Ac-Ala$_n$-LysH$^+$, $n = 5, 10, 15$: Calculated empirical
  ion-mobility cross sections and comparison with experiment
  (Ref. \cite{hudgins-jarrold}). The solid and dashed lines serve only
  as a guide to the eye, and have no physical significance.}
\label{cross-sections}
\end{figure}

The termination-connecting H-bond of the g-1 conformer also leads to
an overall volume of the g-1 conformer that is somewhat smaller than
of the $\alpha$-1, $\alpha$-2, or 3$_{10}$-1 conformers. This is
quantified in Fig. \ref{cross-sections} by way of computed empirical relative
ion-mobility cross sections.\cite{wyttenbach-helden} We show
$\Omega$=($\Omega_{measured}-14.50 n$ {\AA}$^2$) as a function of
peptide chain length $n$, the same expression as used by HRJ
\cite{hudgins-jarrold}. Remarkably, the g-1 conformer for $n$=5 
together with $\alpha$-helical conformers for $n$=10 and 15 (dashed
line) yields exactly the same qualitative behavior as the original
data of HRJ. In contrast, our $\alpha$-1 and $\alpha$-2 conformers
would yield a much shallower drop towards $n$=5, whereas the
3$_{10}$-1 conformer ends up too high.

\begin{table}
  \caption{Energy differences of the four chosen Ac-Ala$_5$-LysH$^+$
  conformers wrt. g-1: Pure DFT-PBE (no vdW),
  DFT-PBE+vdW (PES only), 
  and
  harmonic free energy $F$ at 300~K. All energies in eV.} 
  \label{energies}
\begin{center}\begin{tabular}{ccccc}
    \hline
     & {\scriptsize g-1} &  {\scriptsize $\alpha$-1} & {\scriptsize
     $\alpha$-2 } & {\scriptsize 3$_{10}$-1} \\
    \hline
      DFT-PBE     &  0.0  & 0.04  &  0.08  & 0.04 \\
      DFT-PBE+vdW &  0.0  & 0.09  &  0.11  & 0.19 \\
      $F$(300~K)  &  0.0 & 0.01 &  0.06 & 0.17 \\
    \hline
  \hline
  \end{tabular}
\end{center}
\end{table}

In Table \ref{energies}, we summarize our computed energy
hierarchy. In DFT-PBE+vdW, the g-1 conformer is more stable than 
its closest competitors by 0.1-0.2~eV. On this scale, vdW interactions
are important, as seen by comparing to the pure
DFT-PBE energy hierarchy (no vdW). On the other hand,
finite temperature effects reduce the relative
stability of g-1. The calculated harmonic free energy of g-1 and
$\alpha$-1 at 300~K is almost equal, and $\alpha$-2 is only slightly
($\sim$ 60 meV) less stable; only 3$_{10}$-1 stays
noticeably removed. The expected stability of at least three out of
the four conformers is thus similar. We note in passing that
the hierarchy for other DFT \emph{functionals} (revPBE, or B3LYP at
fixed geometry obtained using the PBE functional) is qualitatively 
similar, as long as a vdW correction is included.\footnote{As a test,
  we verified that the fully relaxed $\alpha$-1 geometry with the
  B3LYP functional is very similar to the fully relaxed PBE
  geometry.}  

\begin{figure*}[ht]
\begin{center}
\includegraphics[width=0.75\textwidth]{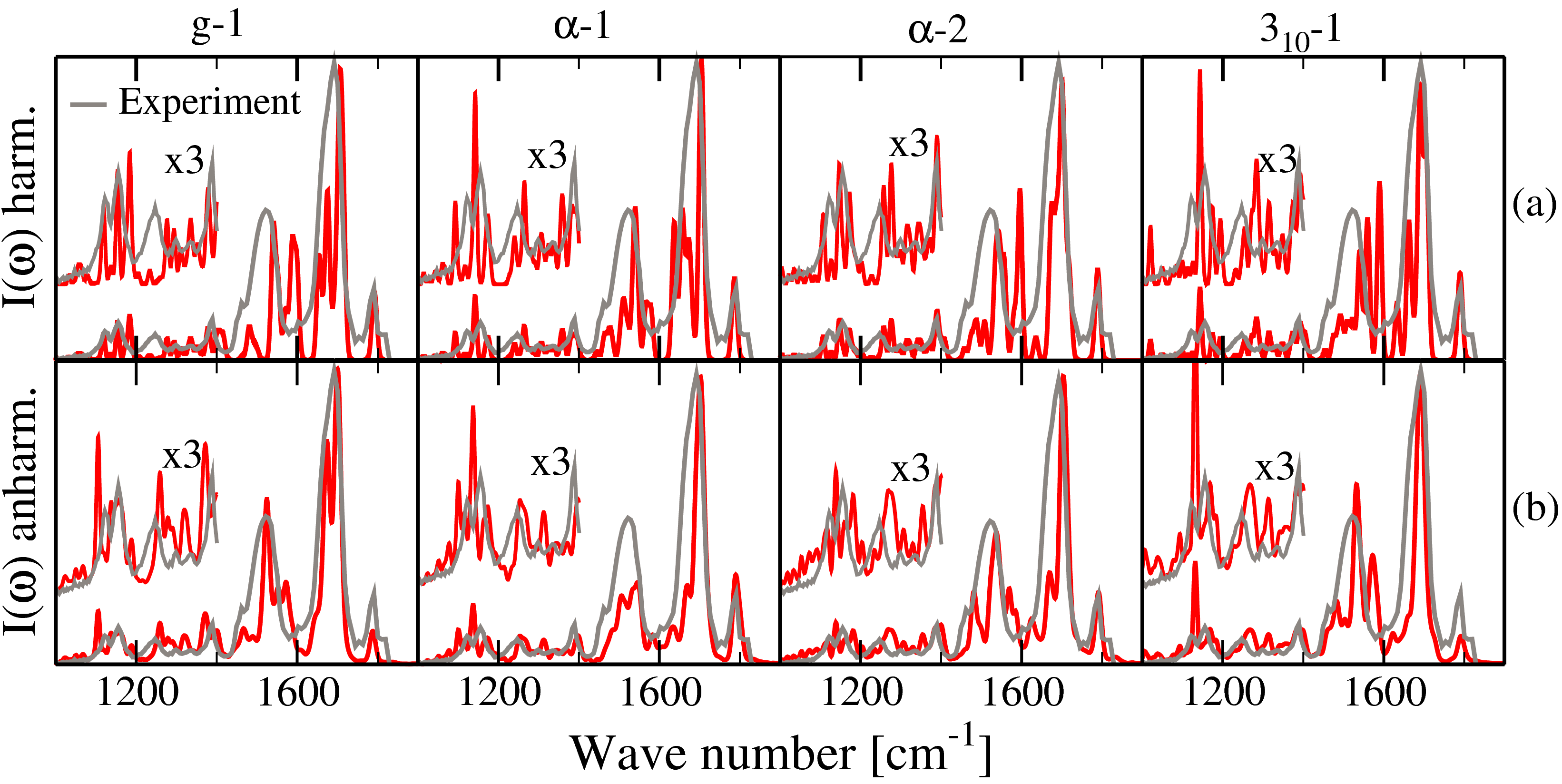}
\vspace*{-5mm}
\end{center}
\caption{Ac-Ala$_5$-LysH$^+$: (a) Theoretical harmonic vibrational
  spectra (red line) for the four chosen conformers compared with experiment
  (gray line); (b) same for \emph{anharmonic} spectra from AIMD
  trajectories. All
  theoretical spectra are shifted rigidly so that the free CO peak ($\sim1790$cm$^{-1}$) is aligned.} 
\label{exp-harm-anharm}
\end{figure*}

Finally, Fig.\ref{exp-harm-anharm} shows computed vibrational
spectra for all four conformers compared to
experiment. Again, we align the localized free C-O vibration peak to experiment by a
rigid shift and scale the intensities to match the height of the Amide-I peak. No further scaling factors are 
employed. In panel (a), spectra calculated in the harmonic
approximation are shown. While there is an overall qualitative
similarity of measured and computed spectra for \emph{all} four
conformers, it is also clear that none of them fit entirely---peak
shifts and incorrect relative intensities (especially those obscuring
the gap between Amide-I and -II) abound. This situation changes when
spectra computed from \emph{ab initio} molecular dynamics and the
dipole-dipole autocorrelation are considered [panel (b)]. For g-1,
$\alpha$-1, and $\alpha$-2, nearly perfect matches to
experiment are obtained: The relative positions of Amide-I and
Amide-II are almost exact, the interfering peaks in the gap decrease 
in intensity, and even the fine structure towards lower wave numbers
is well reproduced. As one example, consider the experimental peak at
$\approx$1250~cm$^{-1}$ compared to the g-1 conformer. It coincides
with a minimum of the harmonic spectrum, while a theoretical peak lies much
closer in the anharmonic case. Consistent with the free energy, it is thus
possible and plausible that all three conformers contribute to the
experimentally observed signal from the ions, which were held in the ion trap at room temperature. In contrast,
the features of the theoretical 3$_{10}$-1 conformer do not match
quite as well. If 3$_{10}$-1 is present in the ion trap at all, then
certainly with a much smaller fraction than g-1, $\alpha$-1, and
$\alpha$-2.  

In summary, we demonstrate a \emph{quantitative} structure
prediction for Ac-Ala$_n$-LysH$^+$ ($n$=5,10,15), with strong support
by the good agreement between calculated and measured vibrational
spectra. Our calculations provide a direct confirmation for the
proposed $\alpha$-helical nature of Ac-Ala$_n$-LysH$^+$
($n$=10,15), while the \emph{lowest energy} conformer of
the ``classic'' HRJ series, $n$=5, is indeed \emph{not} a simple
helix. Importantly, finite-temperature free-energy effects still
render $\alpha$-helical $n$=5 conformers possible and even rather
probable in an experimental molecular beam.

\bibliography{polyalanines}

\end{document}